\documentclass[english]{cccconf}
\usepackage{amsmath,amsfonts}
\usepackage{amsthm}
\usepackage[comma,numbers,square,sort&compress]{natbib}
\usepackage{epstopdf}
\usepackage{algorithm}
\usepackage{algorithmic}
\usepackage{array}
\usepackage{verbatim}
\usepackage{graphicx}
\usepackage{xcolor}
\usepackage{subfig}
\usepackage{stfloats}
\usepackage{pgfplots}
\usepackage{enumitem}  
\usepackage{marvosym}  

\begin{document}

\title{Resource Allocation and Pricing for Blockchain-enabled Metaverse: A Stackelberg Game Approach}

\author{Zhanpeng Zhu\aref{zjnu},
        Feilong Lin\aref{zjnu},
        Changbing Tang\aref{zjnu},
        Zhongyu Chen\textsuperscript{\Letter}\aref{zgvtuc,zjnu}
        }


\affiliation[zjnu]{School of Computer Science and Technology, Zhejiang Normal University, Jinhua 321004, P.~R.~China
        \email{\{1656695753, bruce\_lin, tangcb\}@zjnu.edu.cn}}

\affiliation[zgvtuc]{School of Information, Zhejiang Guangsha Vocational and Technical University of Construction, Jinhua 322100, P.~R.~China
\email{czy@zjnu.edu.cn}
}


\maketitle

\begin{abstract}
  As the next-generation Internet paradigm, the metaverse can provide users with immersive physical-virtual experiences without spatial limitations. However, there are various concerns to be overcome, such as resource allocation, resource pricing, and transaction security issues. To address the above challenges, we integrate blockchain technology into the metaverse to manage and automate complex interactions effectively and securely utilizing the advantages of blockchain. With the objective of promoting the Quality of Experience (QoE), Metaverse Service Users (MSUs) purchase rendering and bandwidth resources from the Metaverse Service Provider (MSP) to access low-latency and high-quality immersive services. The MSP maximizes the profit by controlling the unit prices of resources. In this paper, we model the interaction between the MSP and MSUs as a Stackelberg game, in which the MSP acts as the leader and MSUs are followers. The existence of Stackelberg equilibrium is analyzed and proved mathematically. Besides, we propose an efficient greedy-and-search-based resource allocation and pricing algorithm (GSRAP) to solve the Stackelberg equilibrium (SE) point. Finally, we conduct extensive simulations to verify the effectiveness and efficiency of our designs. The experiment results show that our algorithm outperforms the baseline scheme in terms of improving the MSP's profit and convergence speed.
\end{abstract}



\section{Introduction}
\vspace{-2mm}
\par The metaverse has attracted immense public attention \quad\quad recently due to its enormous potential in various application scenarios. MSUs can work, play, and interact socially by digital avatars in the three-dimensional (3D) virtual world without spatial limits through existing advanced technologies, such as Blockchain, Virtual Reality (VR)/Augmented Reality (AR) technologies, Artificial Intelligence (AI) and so on \cite{roadmap-metaverse-2023}. Equipped with dedicated Metaverse devices, such as Head-Mounted Displays (HMDs) and depth-sensing \quad\quad cameras, MSUs can access immersive services provided by the MSP \cite{A-Survey-Metaverse-Fundamentals-Security-Privacy-2023}. By renting the Base Stations (BSs) with \quad\quad\quad available resources (e.g., bandwidth, CPU, and GPU), the MSP can perform various tasks including remote rendering and data transmission for MSUs to obtain profits.

\par Although the metaverse holds significant economic and commercial potential, there are lots of challenges in the \quad\quad construction and implementation process. One of them is resource allocation. Different metaverse applications has \quad distinct resource demands. For instance, when MSUs visit a virtual museum (e.g., Decentraland), a large number of rendering resources are required because of real-time rendering of high-precision 3D artwork models. Clearly, the corresponding bandwidth demand will be high. On the contrary, the virtual meeting (e.g., Zoom) has a low demand for rendering resources but a high demand for bandwidth. The irrational allocation will lead to low utilities of the MSP and MSUs. Some existing works have investigated several typical allocation strategies. For example, Y. Jiang $et\;al.\;$\cite{QoE-Analysis-2023} designed a Joint Resource Allocation and Metaverse service Selection (JRAMS) scheme, in which computing resources of each BS would be evenly allocated to the served MSUs. We can find that some MSUs' low-latency demands cannot be satisfied in this situation even if their budgets are enough. W. C. Ng $et\;al.\;$\cite{Unified-Resource-Allocation-2022} proposed a stochastic optimal resource allocation scheme (SORAS) and adopted the stochastic integer programming method to minimize the total cost of the virtual service provider. The authors in $et\;al.\;$\cite{Bandwidth-Allocation-Pricing-2024} designed a resource allocation scheme based on dynamic programming to maximize the revenue of the server while ensuring the cost of each user. Based on the existing works, how to optimize the resource allocation scheme to further improve the MSP's profit under multiple constraints is still worth more study.

\par There is another challenging problem to be solved, i.e., the pricing mechanism of resources. Obviously, the irrational pricing strategy may bring a series of problems, such as high latency and slow response. There are some researches that utilize several typical methods (e.g., game theory, auction mechanism) to investigate the pricing mechanism in the metaverse. For example, M. Xu $et\;al.\;$\cite{Learning-Based-Incentive-Mechanism-2022} developed a double Dutch auction mechanism for the optimal resource \quad\quad pricing in the metaverse trading market. X. Huang $et\;al.\;$\cite{Joint-User-Association-and-Resource-Pricing-2022} formulated the interaction between MSPs and MSUs as a Stackelberg game, and then utilized distributed and centralized methods to find the optimal selling price of bandwidth resources. Although some existing works have made certain progress, they only focus on the pricing mechanism of either rendering resources or bandwidth resources. For instance, the authors in \cite{Joint-User-Association-and-Resource-Pricing-2022} ignored the pricing of rendering resources. Actually, it is necessary to design a joint pricing mechanism for both rendering and bandwidth resources, which is considered in our work.

\par Due to the virtual and decentralized characteristics of the metaverse, effective transaction process management and trustful identity authentication among entities are more difficult to guarantee compared to common applications \cite{Credit-Investigation-2023}. There exist various risks in the traditional centralized scheme. For example, transaction records among entities may be tampered with or forged. Moreover, attackers may impersonate legitimate MSUs to maliciously raise resource prices if the central authorization node fails or is compromised. Fortunately, the above problems can be solved by the blockchain technology. In the blockchain-enable metaverse, each MSU can ensure the credibility and uniqueness of its own identity information thanks to the immutability and transparency of blockchain. Additionally, the security and reliability of resource transactions can be guaranteed by the consensus and smart contract mechanism \cite{Revolutionizing-Virtual-Shopping-UAV-2023,MetaChain-2022}. The reason is that the algorithms predefined in the smart contract will be executed automatically without the interference of a trusted third party once specified conditions are met \cite{Smart_Contract-2021}.

\par In a real social metaverse scenario, MSUs achieve \quad\quad immersive virtual experiences by purchasing multiple resources from the MSP. In this paper, we take two types \quad\quad\quad\quad of resources into consideration, specifically rendering and bandwidth resources. We consider a scenario that includes multiple MSUs and BSs equipped with rendering servers. These BSs are managed by one MSP. We model and study the allocation and pricing problem of multiple resources \quad\quad between the MSP and MSUs. Our main contributions are as follows:

\begin{itemize}[topsep=1pt]
  \item A blockchain-enabled metaverse resource allocation and pricing framework is proposed to support secure and efficient resource transactions, in which rational and irrational MSUs are both considered.
  \item We model the interaction between the MSP and MSUs as a Stackelberg game, in which the MSP acts as the leader and MSUs are followers. Moreover, the existence of Stackelberg equilibrium is analyzed and proved mathematically.
  \item We propose an efficient greedy-and-search-based \quad\quad resource allocation and pricing algorithm (GSRAP), which is composed of two layers. In the outer layer, the algorithm solves the optimal solution of resource prices via golden section search. In the inner layer, the algorithm utilizes the greedy strategy to conduct the first resource allocation and then exchange the mapping relationship between BSs and MSUs to improve the MSP's profit further.
  \item We conduct extensive simulations. The results show that our proposed algorithm outperforms the baseline scheme in terms of improving the MSP's profit and \quad convergence speed.
\end{itemize}

\par The remainder of this paper is structured as follows. \quad\quad Section 2 introduces the system model. Section 3 presents and analyzes the Stackelberg game. An effective algorithm is proposed to solve the Stackelberg equilibrium point in Section 4. Extensive simulations are designed and conducted in Section 5. Finally, Section 6 concludes this paper.

\section{System Model}
\par We consider such a social metaverse scenario, there are multiple BSs, a certain amount of MSUs, and the MSP. Let $\mathcal{B}=\{b_1,b_2,...,b_M\}$ denote the set of base stations and $\mathcal{U}=\{u_1,u_2,...,u_N\}$ represent the set of MSUs. Each MSU is equipped with a Head-Mounted Display (HMD) and a depth-sensing camera integrated with wireless communication modules. The HMDs are responsible for collecting head movement data and eye gaze data of MSUs, which can be denoted as $\mathcal{H}=\{H_1,H_2,...,H_i,...,H_n\}$. The depth-sensing cameras collect body movement data of MSUs. The HMD transmits the collected data to the depth-sensing camera via Bluetooth, then the camera forwards the data to the BSs for further processing and corresponding VR video frame rendering. Following that, the HMD downloads the rendered Field of View (FoV) data from the BSs and displays the contents for MSUs.

\subsection{MSP-to-MSUs Transaction Process}
\par The resource transactions between the MSP and MSUs can be managed and automated efficiently through the blockchain. In practice, the MSP will first broadcast the unit resource prices and the number of available resources in terms of a smart contract to the blockchain network \cite{MetaChain-2022}. Then MSUs will generate transactions including budgets and benefit coefficients and send them to the smart contract after signing with their private keys \cite{Pricing-and-Budget-Allocation-2023}. Finally, the smart contract will be enforced automatically once the consensus is reached. After the execution of the smart contract, MSUs obtains resources to enjoy immersive physical-virtual experiences and the MSP gets the payments from MSUs. The transaction results will be permanently recorded in the blockchain, which cannot be altered.

\subsection{MSU Utility Model}
\par Here we assume that the quantity of rendering resources purchased by MSU $u_i$ in one playback period is $x_i^r$. Inspired by \cite{Pricing-and-Budget-Allocation-2023,Cloud-Edge-Computing-2021}, we can evaluate the experience gain obtained by MSU $u_i$ through adopting a logarithmic function as
\begin{equation}
\label{Rir}
R_i^r=\alpha_i\log(1+\mu x_i^r),
\end{equation}

\noindent where $\alpha_i$ is the benefit coefficient of MSU $u_i$ and $\mu$ represents the rendering capacity contained in one unit rendering resource.

\par In a realistic social metaverse scenario, the downlink data transmission has a significantly greater impact on the QoE of MSUs than the uplink. Assuming the amount of bandwidth resources purchased by MSU $u_i$ in one playback period is $x_i^w$ , according to \cite{QoE-Analysis-2023}, we define the Signal-to-Interference-plus-Noise-Ratio (SINR) of the head-mounted display $H_i$ as
\begin{equation}
\label{SINR}
SINR_i=\frac{p_ih_i}{B_0\varphi x_i^w+\epsilon_i^2},
\end{equation}
where $p_i$ is the transmission power of HMD $H_i$, and $B_0$ \quad denotes the noise power spectral density. $h_i=|g_i|^2d_{i,j}^{-\delta}$ \quad\quad represents the channel gain, where $g_i\sim\mathcal{CN}(0,1)$ is the small-scale fading \cite{Demand-Response-2023}, $d_{i,j}$ is the distance between HMD $H_i$ and BS $b_j$, $\delta$ is the path loss exponent. $\epsilon_i^2$ is the interference. Then, we can define the experience gain obtained by MSU $u_i$ from purchasing bandwidth resources as
\begin{equation}
\label{Riw}
\begin{aligned}
R_i^w &= \beta_ix_i^wSINR_i \\
&=\frac{\beta_ix_i^wp_ih_i}{B_0\varphi x_i^w+\epsilon_i^2},
\end{aligned}
\end{equation}
\noindent where $\beta_i$ is the payoff coefficient. Since different MSUs have varying tolerance to latency in VR videos, we assign distinct utility parameters for them.

\par Now we can define the MSU's total utility function as
\begin{equation}
\label{Pi}
P_i = \alpha_i \log(1 + \mu x_i^r) + \frac{\beta_i x_i^w p_i h_i}{B_0 \varphi x_i^w + \epsilon_i^2 }-p_r x_i^r - p_w x_i^w,
\end{equation}
where $p_r$ and $p_w$ are the prices of rendering and bandwidth resources, respectively.

\par Here, we can calculate the first derivatives of the utility function $P_i$ with respect to $x_i^r$ and $x_i^w$ as follows:
\begin{equation}
\label{Pixir111}
\frac{\partial {P_i}}{\partial {x_i^r}}=\frac{\mu \alpha_i}{1+\mu x_i^r}-p_r;\;\frac{\partial {P_i}}{\partial {x_i^w}}=\frac{\beta_i\epsilon_i^2h_ip_i}{(B_0\varphi x_i^w+\epsilon^2)^2}-p_w.
\end{equation}

\par To make the problem more reasonable, $\frac{\partial{P_i}}{\partial{x_i^r}}(0)\ge 0$ and $\frac{\partial{P_i}}{\partial{x_i^w}}(0)\ge 0$ need to be satisfied, i.e., $\{p_r\le \mu \alpha_i,\;p_w\le\frac{\beta_ih_ip_i}{\epsilon_i^2},\;\forall u_i\in\mathcal{U}\}$ should be held, otherwise MSUs will never buy resources from the MSP. Clearly, the MSP will set the prices satisfying the following inequalities:
\begin{equation}
\label{prpw_limit}
p_r\le p_r^{max};\;p_w\le p_w^{max},
\end{equation}
\noindent where $p_r^{max}=\min\{\mu \alpha_i\;|\;\forall u_i\in \mathcal{U}\} u_i\in\mathcal{U}\}$ and $p_w^{max}=\min\{\frac{\beta_ih_ip_i}{\epsilon_i^2}\;|\;\forall u_i\in\mathcal{U}\}$.

\subsection{MSP Utility Model}
\par The MSP's revenue comes from MSUs' payments for rendering and bandwidth resources. We can calculate the total revenue of the MSP as
\begin{equation}
\label{RMSP}
R^{MSP}=\sum_{i=1}^Np_rx_i^r+p_wx_w.
\end{equation}

\par Note that the MSP needs to pay the necessary cost to maintain multiple BSs. We define the cost of the MSP for serving MSU $u_i$ as
\begin{equation}
\label{RMSP}
cost_i=\sum_{j=1}^Mb_{ij}(\xi_{ij}^rx_i^r+\xi_{ij}^wx_i^w),
\end{equation}
where $\xi_{ij}^r$ and $\xi_{ij}^w$ are the costs required by BS $b_j$ to provide unit rendering and bandwidth resource for MSU $u_i$, respectively. $b_{ij}$ is a binary variable, that is, $b_{ij}=\{0,1\}$, which indicates that MSU $u_i$ is assigned to BS $b_j$ to obtain services if $b_{ij}=1$. Here we assume $\sum_{j=1}^Mb_{ij}=1$ is held.

\par Then, the utility function of the MSP can be calculated by
\begin{equation}
\label{utility_function_MSP}
U^{MSP}=\sum_{i=1}^Np_rx_i^r+p_wx_i^w-cost_i.
\end{equation}

\par Apparently, $p_r\ge p_r^{min}$ and $p_w\ge p_w^{min}$ should be satisfied for the MSP, where $p_r^{min}=\max\{\xi_{ij}^r\;|\;\forall u_i\in\mathcal{U},\forall b_j\in\mathcal{B}\}$ and $p_w^{min}=\max\{\xi_{ij}^w\;|\;\forall u_i\in\mathcal{U},\forall b_j\in\mathcal{B}\}$. Combining the MSP's upper bound on the selling prices in \eqref{prpw_limit}, we can get the feasible price ranges of the MSP are
\begin{equation}
\label{prpw_feasible_interval}
p_r^{min}\le p_r\le p_r^{max};\;p_w^{min}\le p_w\le p_w^{max}.
\end{equation}

\section{Stackelberg Game between the MSP and MSUs}
\par We formulate the interaction process between the MSP and MSUs as a Stackelberg game, in which the MSP acts as the leader and MSUs are followers. The game can be divided into two subgames: MSP subgame and MSU subgame. Specifically, the MSP determines the unit prices of resources according to the first-move advantage and decides which BS to provide service for MSUs to maximize its profit in MSP subgame. Each MSU decides the amount of rendering and bandwidth resources to purchase for maximizing their utility in MSU subgame.

\subsection{Problem Formulation}
\par According to the previous analysis, here we give a \quad\quad detailed definition of the optimization problem of MSU $u_i$ in the MSU subgame as follows:
\smallskip
\begin{equation}
\label{MSU_subgame_problem}
    \max_{x_i^r, x_i^w} \, P_i
\end{equation}
\begin{equation}
\label{MSU_subgame_problem_limit_1}
\text{s.t.} \quad  x_i^r \ge 0, x_i^w \ge 0,
\end{equation}
\begin{equation}
\label{MSU_subgame_problem_limit_2}
\quad\quad\quad\quad p_r x_i^r+p_w x_i^w \le B_i. 
\end{equation}

\par The constraint \eqref{MSU_subgame_problem_limit_2} indicates that the total expenditure on rendering and bandwidth resources cannot exceed $B_i$, where $B_i$ represents the budget of MSU $u_i$.

\par In MSP subgame, the MSP maximizes the profit by joint controlling the pricing strategy and determining resource allocation. Let $\boldsymbol{\mathit{b}}=\{\{b_{ij}\}_{j=1}^M\}_{i=1}^N$ be the decision variable for the MSP. Then, the optimization problem of the MSP can be defined as
\begin{align}
\label{MSP_subgame_problem}
&\max_{p_r, p_w,\boldsymbol{\mathit{b}}} \, U^{MSP}\\
\text{s.t.} \quad  &p_r^{min}\le p_r\le p_r^{max},\\
&p_w^{min}\le p_w\le p_w^{max},\\
&\sum_{j=1}^Mb_{ij}=1,\;\forall u_i\in\mathcal{U},\\
&b_{ij}=\{0,1\}.\quad
\end{align}
\subsection{Stackelberg Equilibrium Analysis}
In the following, we first prove that Nash equilibrium exists in MSU subgame and derived the optimal amount of rendering and bandwidth resources purchased by a MSU. Then we prove the existence of Stackelberg equilibrium, at which neither the MSP nor any MSU can improve their utility by unilaterally changing the strategy.

\par \textit{Lemma 1:} Nash equilibrium exists in MSU subgame.
\begin{proof}
Apparently, the MSU's utility function $P_i$ defined in \eqref{Pi} is continuous, and the second-order derivatives of the utility function $P_i$ with respect to $p_r$ and $p_w$ can be \quad\quad\quad\quad calculated as
\begin{align}
\frac{\partial^2 P_i}{\partial (x_i^r)^2} = -\frac{\mu^2 \alpha_i}{(1 + \mu x_i^r)^2};\;\frac{\partial^2 P_i}{\partial x_i^r\partial x_i^w}=0,\\
\frac{\partial^2 P_i}{\partial (x_i^w)^2} = -\frac{2 B_0 \varphi \beta_i \varepsilon_i^2 h_i p_i}{(B_0 \varphi x_i^w + \varepsilon_i^2)^3};\;\frac{\partial^2 P_i}{\partial x_i^w\partial x_i^r}=0.
\end{align}

\par Since parameters are positive, we have $\frac{\partial^2 P_i}{\partial (x_i^r)^2}<0$ and $\frac{\partial^2 P_i}{\partial (x_i^w)^2}<0$. Hence, the utility function $P_i$ is strictly concave about $x_i^r$ and $x_i^w$. Based on the Nash existence theorem, Nash equilibrium exists in MSU subgame \cite{Cloud-Edge-Computing-2021}.
\end{proof}
\par In the following, we will derive the solution of MSU \quad\quad subgame for rational and irrational MSUs at a given pricing \quad\quad strategy $(p_r,p_w)$ from the MSP, respectively.
\subsubsection{Strategies for Rational MSUs}
\par It is obvious that the optimization problem of each MSU $u_i$ is a typical convex optimization problem, which can be solved by the Karush$-$Kuhn$-$Tucker (KKT) conditions.

\par Let $\lambda_1$, $\lambda_2$, and $\lambda_3$ be the Lagrange's multipliers, then we can define the Lagrangian function as
\begin{equation}
\label{Li}
L_i = P_i + \lambda_1 \left( B_i - p_w x_i^w - p_r x_i^r \right) + \lambda_2 x_i^r + \lambda_3 x_i^w.
\end{equation}

\par For MSU $u_i$, the KKT conditions are listed as follows:
\begin{align}
& \quad \frac{\partial L_i}{\partial x_i^r} = \frac{\mu \alpha_i}{1 + \mu x_i^r} - p_r - \lambda_1 p_r + \lambda_2 = 0, \label{Li_xir_equal_0} \\
& \quad \frac{\partial L_i}{\partial x_i^w} = \frac{\beta_i h_i p_i \varepsilon_i^2}{(B_0 \varphi x_i^w + \varepsilon_i^2)^2} - p_w - \lambda_1 p_w + \lambda_3 = 0, \label{Li_xiw_equal_0} \\
& \quad \lambda_1 (B_i - p_w x_i^w - p_r x_i^r) = 0, \label{lambda_1_B_i} \\
& \quad \lambda_2 x_i^r = 0, \label{lambda_2_xir} \\
& \quad \lambda_3 x_i^w = 0, \label{lambda_3_xiw} \\
& \quad \lambda_1, \lambda_2, \lambda_3 \geq 0, \label{lambda_123} \\
& \quad B_i - p_w x_i^w - p_r x_i^r \geq 0, \label{B_i_constraint} \\
& \quad x_i^r \geq 0, x_i^w \geq 0. \label{xir_xiw_ge_0}
\end{align}

\par We can find the optimal solution of the proposed optimization problem in MSU subgame in one of the following cases.
\par $(1)\;Case\;1:$  $x_i^r=0,\;x_i^w=0$. In this case, we have $\lambda_1=0$ as $B_i>0$ according to the KKT condition \eqref{lambda_1_B_i}. Substitute it into \eqref{Li_xir_equal_0} and \eqref{Li_xiw_equal_0}, we can get $\lambda_2=p_r-\frac{\mu \alpha_i}{1 + \mu x_i^r}$ and $\lambda_3=p_w-\frac{\beta_i h_i p_i \varepsilon_i^2}{(B_0 \varphi x_i^w + \varepsilon_i^2)^2}$. Since $p_r\le \mu \alpha_i$ and $p_w\le\frac{\beta_ih_ip_i}{\epsilon_i^2}$, we have $\lambda_2\le0$ and $\lambda_3\le0$. To satisfy the KKT conditions, $\lambda_2=0$ and $\lambda_3=0$ should be held. Specifically, we need to check whether $p_r=\mu\alpha_i$ and $p_w=\frac{\beta_ih_ip_i}{\epsilon_i^2}$ are held, if yes, the optimal solution is $x_i^r=0$, $x_i^w=0$.

\par $(2)\;Case\;2:$  $x_i^r=0,\;x_i^w>0$. As for this case, we have $\lambda_3=0$ according to the KKT condition \eqref{lambda_3_xiw}. Substitute it into \eqref{Li_xir_equal_0} and \eqref{Li_xiw_equal_0}, we have
\begin{equation}
\label{eq2.1}
\mu \alpha_i - p_r - \lambda_1 p_r + \lambda_2 = 0,
\end{equation}
\begin{equation}
\label{eq2.2}
\frac{\beta_i h_i p_i \varepsilon_i^2}{(B_0 \varphi x_i^w + \varepsilon_i^2)^2} - p_w - \lambda_1 p_w = 0.
\end{equation}

\par Substitute $x_i^r=0$ into \eqref{lambda_1_B_i}, we have
\begin{equation}
\label{eq2.3}
\lambda_1 (B_i - p_w x_i^w) = 0.
\end{equation}

\par $Case\;2$-$1:$ If $\lambda_1=0$, then we can rewrite \eqref{eq2.1} as $\mu\alpha_i-p_r+\lambda_2=0$, from which we can get $\lambda_2=p_r-\mu\alpha_i$ easily. Because $p_r\le\mu\alpha_i$, we have $\lambda_2\le0$. The KKT condition \eqref{lambda_123} can only be satisfied when $p_r=\mu\alpha_i$. Substitute $\lambda_1=0$ into \eqref{eq2.2}, we can get the expression of $x_i^w$ as follows:
\begin{equation}
\label{eq2.4}
x_i^w = \frac{1}{B_0 \varphi} \left( \sqrt{\frac{\beta_i h_i p_i \varepsilon_i^2}{p_w}} - \varepsilon_i^2 \right).
\end{equation}

\par $Case\;2$-$2:$ If $\lambda_1>0$, then we have $B_i-p_wx_i^w=0$ according to \eqref{eq2.3}. Then we can obtain $x_i^w=\frac{B_i}{p_w}$. Substitute it into \eqref{eq2.2}, we have
\begin{equation}
\label{eqxxxxx}
\lambda_1 = \frac{\beta_i h_i p_i \varepsilon_i^2}{p_w \left( B_0 \varphi \frac{B_i}{p_w} + \varepsilon_i^2 \right)^2} - 1.
\end{equation}

\par By substituting \eqref{eqxxxxx} into \eqref{eq2.1}, we have $\lambda_2=p_r+\lambda_1p_r-\mu\alpha_i$. We need to evaluate whether $\lambda_1>0$ and $\lambda_2\ge0$ are both satisfied simultaneously, if yes, the optimal solution is $x_i^r=0,\;x_i^w=\frac{B_i}{p_w}$.

\par $(3)\;Case\;3:$  $x_i^r>0,\;x_i^w=0$. For this case, we can get $\lambda_2=0$ based on \eqref{lambda_2_xir}. Then we can rewrite the KKT conditions \eqref{Li_xir_equal_0} and \eqref{Li_xiw_equal_0} as
\begin{equation}
\label{eq3.1}
\frac{\mu \alpha_i}{1 + \mu x_i^r} - p_r - \lambda_1 p_r = 0.
\end{equation}
\begin{equation}
\label{eq3.2}
\frac{\beta_i h_i p_i}{\varepsilon_i^2} - p_w - \lambda_1 p_w + \lambda_3 = 0.
\end{equation}

\par $Case\;3$-$1:$ Consider $\lambda_1=0$. Substitute it into \eqref{eq3.1} and \eqref{eq3.2}, we have $x_i^r = \frac{\alpha_i}{p_r} - \frac{1}{\mu}$ and $\lambda_3=p_w-\frac{\beta_i h_i p_i}{\varepsilon_i^2}$. Similar to case 2-1, the KKT condition \eqref{lambda_123} is held only when $p_w=\frac{\beta_i h_i p_i}{\varepsilon_i^2}$, i.e., $\lambda_3=0$. Then we need to check whether the KKT condition \eqref{B_i_constraint} is satisfied, if yes, the optimal solution is $x_i^r=\frac{\alpha_i}{p_r} - \frac{1}{\mu},\;x_i^w=0$.

\par $Case\;3$-$2:$ Consider $\lambda_1>0$, we have $B_i-p_rx_i^r-p_wx_i^w=0$ according to the KKT condition \eqref{lambda_1_B_i}. Since $x_i^w=0$ is held in this case, so we have $x_i^r = \frac{B_i}{p_r}$. Substitute it into \eqref{eq3.1}, we can obtain
\begin{equation}
\label{eqxx}
\lambda_1 = \frac{\mu \alpha_i}{p_r \left(1 + \mu \frac{B_i}{p_r}\right)} - 1.
\end{equation}

\par To satisfy the KKT conditions, we need to check whether $\lambda_1>0$ is held. If yes, we can substitute \eqref{eqxx} into \eqref{eq3.2} to obtain $\lambda_3 = p_w + \lambda_1 p_w - \frac{\beta_i h_i p_i}{\varepsilon_i^2}$. If $\lambda_3\ge0$ is also satisfied, then all the KKT conditions are satisfied, which means the optimal solution is $x_i^r=\frac{B_i}{p_r},\;x_i^w=0$.

\par $(4)\;Case\;4:$  $x_i^r>0,\;x_i^w>0$. According to the KKT conditions \eqref{lambda_2_xir} and \eqref{lambda_3_xiw}, we have $\lambda_2=\lambda_3=0$. Substitute it into \eqref{Li_xir_equal_0} and \eqref{Li_xiw_equal_0}, we can obtain
\begin{equation}
\label{eq4.1}
\frac{\mu \alpha_i}{1 + \mu x_i^r} - p_r - \lambda_1 p_r = 0,
\end{equation}
\begin{equation}
\label{eq4.2}
\frac{\beta_i h_i p_i \varepsilon_i^2}{(B_0 \varphi x_i^w + \varepsilon_i^2)^2} - p_w - \lambda_1 p_w = 0.
\end{equation}

\par $Case\;4$-$1:$ Consider $\lambda_1=0$. Combing \eqref{eq4.1} and \eqref{eq4.2}, we can have
\begin{equation}
\label{eq4.3}
x_i^r=\frac{\alpha_i}{p_r} - \frac{1}{\mu};\;x_i^w=\frac{1}{B_0 \varphi} \left( \sqrt{\frac{\beta_i h_i p_i \varepsilon_i^2}{p_w}} - \varepsilon_i^2 \right).
\end{equation}

\par Subsequently, we have to evaluate whether $B_i - p_w x_i^w - p_r x_i^r \geq 0$ is held. If yes, it is apparent that all the KKT conditions are satisfied, and we can conclude that the optimal solution of MSU $u_i$ can be written as Eq. \eqref{eq4.3}.

\par $Case\;4$-$2:$ Consider $\lambda_1>0$. According to (28), we have $B_i - p_w x_i^w - p_r x_i^r = 0$. Combining \eqref{eq4.1} and \eqref{eq4.2}, we obtain
\begin{equation}
\label{eq4.6}
\frac{\beta_i h_i p_i \varepsilon_i^2}{p_w \left( B_0 \varphi x_i^w + \varepsilon_i^2 \right)^2} = \frac{\mu \alpha_i}{p_r \left( 1 + \mu x_i^r \right)}.
\end{equation}

\par Substitute $x_i^r=\frac{B_i-p_wx_i^w}{p_r}$ into \eqref{eq4.6}, then we can obtain a quadratic equation with respect to $x_i^w$. Discarding the unreasonable negative value, we have
\begin{equation}
\label{eq4.7}
x_i^w = -\frac{\varepsilon_i^2}{B_0 \varphi} - \frac{\varepsilon_i^2 \beta_i h_i p_i}{2 \alpha_i B_0^2 \varphi^2} + \frac{\sqrt{A + B + C + D}}{2 \mu \alpha_i B_0^2 \varphi^2 p_w},
\end{equation}

\noindent where $A=\mu^2 \varepsilon_i^4 \beta_i^2 h_i^2 p_i^2 p_w^2,\;B=4 \mu^2 \alpha_i \varepsilon_i^4 \beta_i h_i p_i B_0 \varphi p_w^2,\;C=4 \mu \alpha_i \varepsilon_i^2 \beta_i h_i p_i B_0^2 \varphi^2 p_r p_w$, $D=4 \mu^2 \alpha_i \varepsilon_i^2 \beta_i h_i p_i B_0^2 \varphi^2 B_i p_w$.

\par According to \eqref{eq4.1}, we can get $\lambda_1=\frac{\mu \alpha_i}{(1 + \mu x_i^r) p_r} - 1$, which need to be checked. If $\lambda_1>0$ is satisfied, then the optimal solution of MSU $u_i$ is $x_i^w=$ \eqref{eq4.7}, $x_i^r =\frac{B_i-p_wx_i^w}{p_r}$.

\par It is obvious that the corresponding optimal solution must be found in one of the four cases for a rational MSU. The rational MSUs will always choose the best purchasing \quad\quad\quad\quad strategy to maximize their individual utility.

\vspace{-2mm}
\subsubsection{Strategies for Irrational MSUs}
\par Due to the lack of game awareness, irrational MSUs \quad \quad\quad often determine the allocation of budget according to their own preferences. It is obvious that they may not be capable to maximize the utility. However, the phenomenon usually\quad appears in reality.

\par We assume that the proportion of a irrational MSU's\quad\quad budget spent on purchasing rendering resources is $\gamma_i$, where $0\le\gamma_i\le1$ is held. Then the amount of rendering and bandwidth resources to be bought by MSU $u_i$ as follows:
\begin{equation}
\label{irrational_user_xir}
x_i^r=\frac{\gamma_iB_i}{p_r};\;x_i^w=\frac{(1-\gamma_i)B_i}{p_w}.
\end{equation}

\par \textit{Theorem 1:} The Stackelberg equilibrium exists in the \quad\quad\quad Stackelberg game between the MSP and MSUs.
\begin{proof}
To prove the existence of the Stackelberg equilibrium, we need to analyze the utility function $U^{MSP}$ \quad\quad\quad\quad defined in \eqref{utility_function_MSP}, which can be rewritten as \eqref{re_utility_function_MSP}. The strategy space of the MSP is $[p_r^{min},p_r^{max}]\times[p_w^{min},p_w^{max}]$, which is a \quad\quad non-empty, closed and convex subset of the euclidean space.
\begin{equation}
\label{re_utility_function_MSP}
U^{MSP}=\sum_{i=1}^N\sum_{j=1}^M b_{ij}(p_rx_i^r+p_wx_i^w-\xi_{ij}^rx_i^r-\xi_{ij}^wx_i^w).
\end{equation}

%

\par We can have the second derivatives of $U^{MSP}$ with respect to $p_r$ and $p_w$ as \eqref{utility_function_MSP_2_daoshu_pr} and \eqref{utility_function_MSP_2_daoshu_pw}. In the following, we first consider the cases except case 4-2.
\begin{equation}
\label{utility_function_MSP_2_daoshu_pr}
\begin{aligned}
\frac{\partial^2 U^{\text{MSP}}}{\partial (p_r)^2} =& \sum_{i=1}^{N} \sum_{j=1}^{M} b_{ij} \times \{2 \frac{\partial x_i^r}{\partial p_r} + \left( p_r - \xi_{ij}^r \right) \times \\
 & \frac{\partial^2 x_i^r}{\partial (p_r)^2}+  \left( p_w - \xi_{ij}^w \right) \frac{\partial^2 x_i^w}{\partial (p_r)^2} \},
\end{aligned}
\end{equation}
\begin{equation}
\label{utility_function_MSP_2_daoshu_pw}
\begin{aligned}
\frac{\partial^2 U^{\text{MSP}}}{\partial (p_w)^2} =&
    \sum_{i=1}^{N} \sum_{j=1}^{M} b_{ij} \{ 2 \frac{\partial x_i^w}{\partial p_w} + \left( p_r - \xi_{ij}^r \right) \times \\
    &\frac{\partial^2 x_i^r}{\partial (p_w)^2}  + \left( p_w - \xi_{ij}^w \right) \frac{\partial^2 x_i^w}{\partial (p_w)^2} \}.
\end{aligned}
\end{equation}

\par We can find that the amount of rendering resource to be purchased by MSU $u_i$ must be one of the following three cases: $(1)x_i^r=0$; $(2)x_i^r = \frac{\alpha_i}{p_r} - \frac{1}{\mu}$; $(3)x_i^r=\frac{B_i}{p_r}$. It is evident that the first-order and second-order derivatives of $x_i^r$ and $x_i^w$ with respect to $p_w$ are 0. We can first calculate the first-order derivative of $x_i^r$ with respect to $p_r$ as $\frac{\partial x_i^r}{\partial p_r}=(1)0;$ $(2)-\frac{\alpha_i}{p_r^2};(3)-\frac{B_i}{p_r^2}$, respectively. Then, the second-order derivative of $x_i^r$ with respect to $p_r$ can be calculated as $\frac{\partial^2 x_i^r}{\partial (p_r)^2}=(1)0;(2)\frac{2 \alpha_i}{p_r^3};(3)\frac{2 B_i}{p_r^3}$, respectively.

\par Similarly, we can calculate the first-order derivative of $x_i^w$ with respect to $p_w$ as $\frac{\partial x_i^w}{\partial p_w}=(a)0;(b)-\frac{B_i}{p_w^2};(c)-\frac{\beta_i h_i p_i \varepsilon_i^2}{2 B_0 \varphi \sqrt{\beta_i h_i p_i \varepsilon_i^2 p_w^3}}$, respectively. Then, the second-order derivative of $x_i^w$ with respect to $p_w$ can be calculated as $\frac{\partial^2 x_i^w}{\partial (p_w)^2}=(a)0;(b)\frac{2 B_i}{p_w^3};(c)\frac{3 \beta_i^2 h_i^2 p_i^2 \varepsilon_i^4 p_w^2}{4 B_0 \varphi \left( \beta_i h_i p_i \varepsilon_i^2 p_w^3 \right)^{3/2}}$.

\par Combing the above analysis, we have $\frac{\partial^2 U^{\text{MSP}}}{\partial (p_r)^2}\le0$ and $\frac{\partial^2 U^{\text{MSP}}}{\partial (p_w)^2}\le0$.

\par Subsequently, we consider the case 4-2. By calculation and arrangement, we first obtain the second-order derivative of the utility function $U^{MSP}$ with respect to $p_r$ as
\begin{equation}
\label{case4-2_utility_daoshu_2_MSP_pr}
\frac{\partial^2 U^{\text{MSP}}}{\partial (p_r)^2} = -\sum_{i=1}^{N} \sum_{j=1}^{M} b_{ij} \left( \xi_{ij}^r \frac{\partial^2 x_i^r}{\partial (p_r)^2} + \xi_{ij}^w \frac{\partial^2 x_i^w}{\partial (p_r)^2} \right).
\end{equation}

\par Combing $\frac{\partial x_i^r}{\partial p_r}$, $\frac{\partial^2 x_i^r}{\partial (p_r)^2}$, $\frac{\partial x_i^w}{\partial p_w}$, and $\frac{\partial^2 x_i^w}{\partial (p_w)^2}$, we can rewrite \eqref{case4-2_utility_daoshu_2_MSP_pr} as \eqref{utility_function_MSP_2_daoshu_pr_case4-2}, where $\pi_i=\beta_i \varepsilon_i^2 h_i p_i p_w$ and $E= A + B + C + D$. If $\;\xi_{ij}^r p_w - \xi_{ij}^w p_r=0$, then $\frac{\partial^2 U^{\text{MSP}}}{\partial (p_r)^2}<0$ is satisfied. Similarly, we can have $\frac{\partial^2 U^{\text{MSP}}}{\partial (p_w)^2}<0$. Thus, the utility function of the MSP is quasi-concave in this case \cite{Joint-User-Association-and-Resource-Pricing-2022}.
\begin{equation}
\label{utility_function_MSP_2_daoshu_pr_case4-2}
\begin{split}
\frac{\partial^2 U^{\text{MSP}}}{\partial (p_r)^2} = &-\sum_{i=1}^{N} \sum_{j=1}^{M} \frac{2 b_{i,j}}{p_r^2 E^{3/2}}   \xi_{ij}^r\{ ( x_i^r E^{3/2} +  \pi_i E ) \\
&+\left( \mu \alpha_i B_0^2 \varphi^2 \pi_i^2 p_w p_r \right) \left( \xi_{ij}^r - \xi_{ij}^w p_r \right) \}.
\end{split}
\end{equation}

\par For irrational MSUs, we have $\frac{\partial x_i^r}{\partial p_w}=\frac{\partial ^2x_i^r}{\partial (p_w)^2}=0$ and $\frac{\partial x_i^w}{\partial p_r}=\frac{\partial ^2x_i^w}{\partial (p_r)^2}=0$. The second-order derivatives of $x_i^r,x_i^w$ about $p_r$ and $p_w$ are
\begin{align}
\frac{\partial x_i^r}{\partial p_r}=-\frac{\gamma_iB_i}{(p_r)^2}&;\;\frac{\partial ^2x_i^r}{\partial (p_r)^2}=2\frac{\gamma_iB_i}{(p_r)^3},\\
\frac{\partial x_i^w}{\partial p_w}=-\frac{(1-\gamma_i)B_i}{(p_w)^2}&;\;\frac{\partial ^2x_i^w}{\partial (p_w)^2}=2\frac{(1-\gamma_i)B_i}{(p_w)^3}.
\end{align}

\par By substituting them into \eqref{utility_function_MSP_2_daoshu_pr} and \eqref{utility_function_MSP_2_daoshu_pw}, we have $\frac{\partial^2 U^{\text{MSP}}}{\partial (p_r)^2}\le0$ and $\frac{\partial^2 U^{\text{MSP}}}{\partial (p_w)^2}\le0$.

\par According to the above overall analysis, $\frac{\partial^2 U^{\text{MSP}}}{\partial (p_r)^2}\le 0$ and $\frac{\partial^2 U^{\text{MSP}}}{\partial (p_w)^2}\le 0$ are satisfied based on the responses of all MSUs to the pricing strategy from the MSP. Therefore, we have proven that the Stackelberg equilibrium exists.
\end{proof}

\section{Resource Allocation and Pricing Algorithm}
\par It is apparent that the optimization problem of the MSP's profit is a typical mixed-integer programming problem. To solve it, we propose an efficient greedy-and-search-based resource allocation and pricing algorithm (GSRAP), which needs to overcome two subproblems. The first subproblem is how to allocate BSs to process MSUs' resource requests\quad under a given pricing scheme. Another one is how to find the optimal unit prices of resources quickly. For the first subproblem, we design an greedy-and-exchange-based resource allocation algorithm shown in Algorithm \ref{algorithm-resource-allocation}.

\subsection{Resource Allocation Algorithm}
\par We define two resource request lists $L_{req}^r=\{ {x_1^r}^*,{x_2^r}^*,...,{x_i^r}^*,...,{x_N^r}^*\}$ and $L_{req}^w=\{ {x_1^w}^*,{x_2^w}^*,...,{x_i^w}^*,...,{x_N^w}^*\}$, where each ${x_i^r}^*$ and ${x_i^w}^*$ represents the optimal purchasing decision of MSU $u_i$. Then we construct a utility matrix $G_u$ shown in \eqref{Gu_matrix}. We calculate each element $u_{ij}\in G_u$ through $u_{ij} = p_r{x_i^r}^* + p_w{x_i^w}^* - cost_i$. Next, we define an auxiliary matrix $G_v=\{\{v_{ij}\}_{j=1}^M\}_{i=1}^N$ with the same structure as $G_u$. Each $v_{ij}\in G_v$ is calculated by $v_{ij}=a_1(p_r-\xi_{ij}^r)+a_2(p_w-\xi_{ij}^w)$ where $a_1$ and $a_2$ are weight parameters, which is different from \cite{Cloud-Edge-Computing-2021}.
\begin{equation}
\label{Gu_matrix}
G_u = \begin{pmatrix}
u_{11} & \cdots & u_{1j} & u_{1M} \\
\vdots & \ddots & \vdots & \vdots \\
u_{N1} & \cdots & u_{Nj} & u_{NM}
\end{pmatrix}.
\end{equation}

\par Besides, we define a MSU set $\mathcal{U}'=\{u_1',u_2',...,u_i',...,u_K'\}$, where $|\mathcal{U}'|\le N$. Each element $u_i'\in \mathcal{U}'$ is the MSU waiting to be served by a certain BS. We use $C^r=\{c_1^r,c_2^r,...,c_j^r,...,c_M^r\}$ and $C^w=\{c_1^w,c_2^w,...,c_j^w,...,c_M^w\}$ to record the available number of rendering and bandwidth resources of different BSs, respectively.

\begin{algorithm}
\caption{Greedy-and-Exchange Based Allocation}
\label{algorithm-resource-allocation}
\begin{algorithmic}[1] 
\REQUIRE $p_r,\;p_w,\;\mathcal{U},\;\mathcal{B},\;C^r,\;C^w,\;G_u,\;G_u',\;G_v,\;L_{req}^r,\;L_{req}^w$.
\ENSURE $\boldsymbol{\mathit{b}},\;U^{MSP}$.
\STATE Initialize: $\mathcal{U}' \gets \mathcal{U},U^{MSP} \gets 0$;
\WHILE{$\mathcal{U}'\neq\emptyset$}
    \IF{$u_i' \in \mathcal{U}'$}
        \STATE Select $u_i'$ and $b_j$ with $u_{ij}$ having the largest value in $G_v$;
        \IF{$c_j^r\ge {x_i^r}^*$ and $c_j^w\ge {x_i^w}^*$}
            \STATE update $\boldsymbol{\mathit{b}}$, $C^r$, $C^w$, $\mathcal{U}'$, $U^{MSP}$;
        \ELSE
            \STATE $u_{ij} \gets 0$;
        \ENDIF
    \ENDIF
\ENDWHILE

\STATE $\mathcal{U}' \gets \mathcal{U}$;

\WHILE{$\mathcal{U}'\neq \emptyset$}
    \STATE Select a MSU $u_i'$ from $\mathcal{U}'$, $tag \gets 0$;
    \FOR{each $u_{i'}\in \mathcal{U}$}
        \IF{$\Phi(i,j,i',j')>0$ and $c_{j'}^r+{x_{i'}^r}^*\ge {x_i^r}^*$ and $c_{j'}^w+{x_{i'}^w}^*\ge {x_i^w}^*$ and $c_{j}^w+{x_{i}^w}^*\ge {x_{i'}^w}^*$ and $c_{j}^r+{x_{i}^r}^*\ge {x_{i'}^r}^*$}
            \STATE update $\boldsymbol{\mathit{b}}$, $C^r$, $C^w$, $U^{MSP}$, $tag \gets 1$;
        \ENDIF
    \ENDFOR
    \FOR{each $b_k\in \mathcal{B}$}
        \IF{$\phi(i,j,k)>0$ and $c_{k}^r \ge {x_{i}^r}^*$ and $c_{k}^w \ge {x_{i}^w}^*$}
            \STATE update $\boldsymbol{\mathit{b}}$, $C^r$, $C^w$, $U^{MSP}$, $tag \gets 1$;
        \ENDIF
    \ENDFOR
    \IF{$tag = 0$}
        \STATE $\mathcal{U}' \gets \mathcal{U}'-u_i'$;
    \ENDIF
\ENDWHILE

\RETURN $\boldsymbol{\mathit{b}}$, $U^{MSP}$.
\end{algorithmic}
\end{algorithm}

\par The algorithm can be divided into three stages. In the first stage, we attempt to allocate resources to MSUs leveraging the greedy strategy (lines $2\!-\!11$). In each iteration, the $v_{ij}$ with the largest value in matrix $G_v$ will be chosen by the algorithm, and then the algorithm decides whether to assign BS $b_j$ to MSU $u_i$. The process continues until every MSU has been matched to one BS.

\par Subsequently, the algorithm tries to promote the MSP's profit through exchanging the mapping relationship between BSs and MSUs. We assume that MSU $u_i$ and $u_{i'}$ ($i\ne i'$) are matched to BS $b_j$ and $b_{j'}$ at the beginning, respectively. We define $\Phi(i,j,i',j')$ in \eqref{Phi} to represent the possible extra profit. If $\Phi(i,j,i',j')>0$, then the exchange process will be executed.
\begin{equation}
\label{Phi}
\Phi(i,j,i',j')=(u_{ij'}+u_{i'j})-(u_{ij}+u_{i'j'})
\end{equation}

\par The MSP's profit may be promoted further via another \quad \quad exchange strategy during the third stage, as illustrated in lines $20\!-\!24$. Specifically, when a MSU is assigned from one BS to another BS, the cost for serving the MSU is possible to be reduced. Here we define $\phi(i,j,k)$ in \eqref{phi} to evaluate the possible extra profit of assigning $u_i$ from $b_j$ to $b_k$. If $\phi(i,j,k)>0$, then the exchange will be performed.
\begin{equation}
\label{phi}
\phi(i,j,k)=u_{ik}-u_{ij}
\end{equation}

\par Note that Algorithm \ref{algorithm-resource-allocation} proceeds iteratively and finishes\quad until the total profit of the MSP no longer changes. The results will be permanently stored in blockchain.

\subsection{GSRAP Algorithm}
\par In order to solve the SE point of our formulated game\quad\quad effectively, we adopt an improved golden section search \quad\quad\quad strategy in GSRAP to find the optimal unit prices of \quad\quad\quad\quad resources. The pseudo-code of algorithm GSRAP is shown in Algorithm \ref{algorithm-GSRAP}.

\par The algorithm narrows the search ranges of the unit prices iteratively with the golden ratio. Take the update of the bandwidth resource price $p_w$ as an example. First, the initial search range $[p_w^l,p_w^r]$ is set as $[p_w^{min},p_w^{max}]$. Then the algorithm will decide whether the optimal unit bandwidth price falls into $[p_w^l,p_w^l+0.618*(p_w^r-p_w^l)]$ or $[p_w^l+0.382*(p_w^r-p_w^l),p_w^r]$ by calculation and comparison (lines $6\!-\!11$). We improve the algorithm by adding a random value to the bounds of the interval after updating the search range each time, which allows it to adapt to complex situations better. For example, if $p_w^r$ has just updated via $p_w^r={p_w^r}'$, the algorithm will first generate a random number $\omega$ from the interval $[-\kappa(p_w^r-p_w^l),\kappa(p_w^r-p_w^l)]$ where $\kappa$ is a small positive parameter, and then execute $p_w^r=p_w^r+\omega$.

\par Algorithm \ref{algorithm-GSRAP} continues until the intervals of the search ranges related to rendering and bandwidth resource prices are both less than our predefined threshold $\delta$.

\setlength{\textfloatsep}{5pt}  
\begin{algorithm}
\caption{GSRAP} \label{algorithm-GSRAP}
\begin{algorithmic}[1] 
\REQUIRE $p_r^{min}$, $p_r^{max}$, $p_w^{min}$, $p_w^{max}$, $\mathcal{U}$, $\mathcal{B}$, $G_u$, $G_v$.
\ENSURE $\boldsymbol{\mathit{b}}$, $U^{MSP}$.
\STATE $p_r^l \gets p_r^{min}$, $p_r^r \gets p_r^{max}$, ${p_r^l}' \gets p_r^{min}$, ${p_r^r}' \gets p_r^{max}$;
\STATE $p_w^l \gets p_w^{min}$, $p_w^r \gets p_w^{max}$, ${p_w^l}' \gets p_w^{min}$, ${p_w^r}' \gets p_w^{max}$;
\STATE $p_r \gets p_r^{min}$, $p_w \gets p_w^{min}$, $U^{MSP} \gets 0$, $\delta \gets 0.001$;
\WHILE{$p_w^r - p_w^l \ge \delta$ and $p_r^r - p_r^l \ge \delta$}
    \STATE \text{// Optimize the price of bandwidth resource}
    \STATE $p_w \gets {p_w^l}'$, $(val_1,\boldsymbol{\mathit{b_1}}) \gets Algorithm\; 1(p_r,p_w)$;
    \STATE $p_w \gets {p_w^r}'$, $(val_2,\boldsymbol{\mathit{b_2}}) \gets Algorithm\; 1(p_r,p_w)$;
    \IF{$val_1>val_2$}
        \STATE $U^{MSP} \gets val_1$, $\boldsymbol{\mathit{b}} \gets \boldsymbol{\mathit{b_1}}$, update $p_w^r$;
    \ELSE
        \STATE $U^{MSP} \gets val_2$, $\boldsymbol{\mathit{b}} \gets \boldsymbol{\mathit{b_2}}$, update $p_w^l$;
    \ENDIF
    \STATE ${p_w^l}' \gets p_w^l+0.382*(p_w^r-p_w^l)$;
    \STATE ${p_w^r}' \gets p_w^l+0.618*(p_w^r-p_w^l)$;

    \STATE \text{// Optimize the price of rendering resource}
    \STATE $p_r \gets {p_r^l}'$, $(val_3,\boldsymbol{\mathit{b_3}}) \gets Algorithm\; 1(p_r,p_w)$;
    \STATE $p_r \gets {p_r^r}'$, $(val_4,\boldsymbol{\mathit{b_4}}) \gets Algorithm\; 1(p_r,p_w)$;
    \IF{$val_3>val_4$}
        \STATE $U^{MSP} \gets val_3$, $\boldsymbol{\mathit{b}} \gets \boldsymbol{\mathit{b_3}}$, update $p_r^r$;
    \ELSE
        \STATE $U^{MSP} \gets val_4$, $\boldsymbol{\mathit{b}} \gets \boldsymbol{\mathit{b_4}}$, update $p_r^l$;
    \ENDIF
    \STATE ${p_r^l}' \gets p_r^l+0.382*(p_r^r-p_r^l)$;
    \STATE ${p_r^r}' \gets p_r^l+0.618*(p_r^r-p_r^l)$;

\ENDWHILE

\STATE $p_r \gets 0.5*(p_r^l+p_r^r)$, $p_w \gets 0.5*(p_w^l+p_w^r)$;
\RETURN $\boldsymbol{\mathit{b}}$, $p_r$, $p_w$, $U^{MSP}$.
\end{algorithmic}
\end{algorithm}

\section{Numerical Results}
\par In this section, extensive simulations are conducted to validate our designs. First, we perform numerical simulations to prove the existence of the Stackelberg equilibrium. Second, we evaluate the sensitivity of certain parameters. Finally, we compare the proposed algorithms with the baseline scheme.

\subsection{Simulation Setup}
\par In the simulation experiments, we assume the benefit coefficients of MSUs $\alpha_i,\beta_i$ are randomly initialized from $[30,35]$ and $[10,15]$, respectively. The budgets of MSUs are chosen from $[30,100]$. The number of MSU in the metaverse scenario is set to be $100$. To simulate the realistic metaverse service using situation, we set the proportion of rational and irrational MSUs to be $80\%$ and $20\%$, respectively. For each irrational MSU, the proportion of budget used to buy rendering resources $\gamma_i$ is randomly generated from $[0,1]$. In GSRAP, we set $\kappa$ to be $0.1$. The weight parameters $a_1$ and $a_2$  are set to be $0.5$ and $0.5$, respectively. 

\subsection{Stackelberg Equilibrium}
\par
\begin{figure*}[htbp]  
    \centering
    \begin{minipage}{0.24\textwidth}  
        \centering
        \includegraphics[width=\textwidth,height=4.5cm]{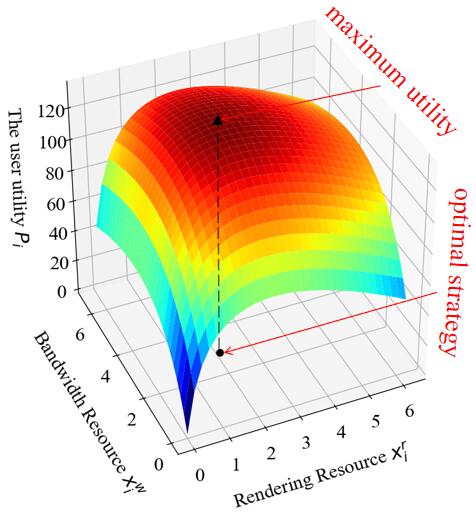}  
        \caption{The optimal strategy of MSU $u_i$.}
        \label{fig-user_utility_xir_xiw}
    \end{minipage} \hfill  
    \begin{minipage}{0.24\textwidth}  
        \centering
        \includegraphics[width=\textwidth,height=4.5cm]{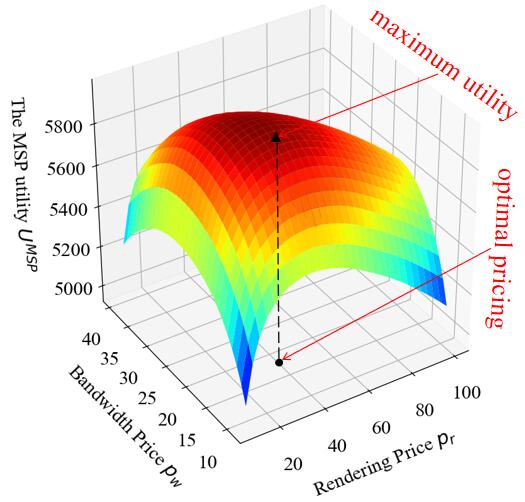}  
        \caption{The optimal strategy of the MSP.}
        \label{fig-MSP_utility_pr_pw}
    \end{minipage} \hfill  
    \begin{minipage}{0.24\textwidth}  
        \centering
        \includegraphics[width=\textwidth,height=4.5cm]{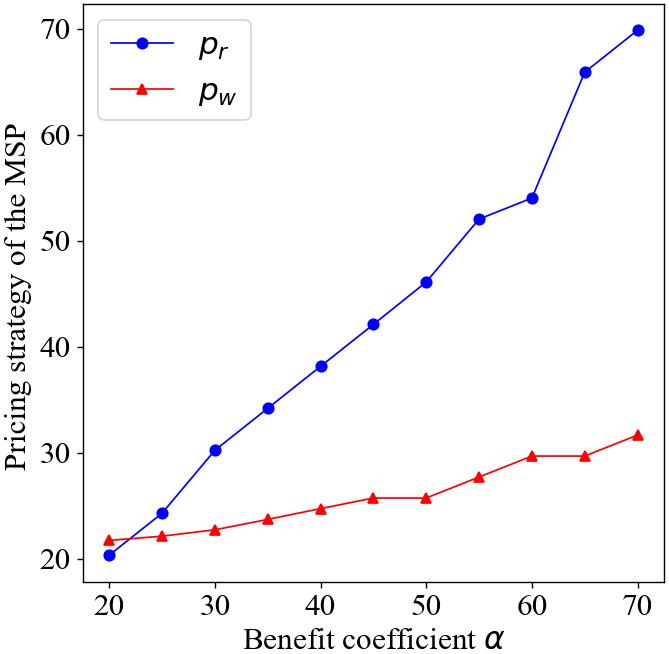}  
        \caption{The effect of $\alpha$ on \quad\quad unit prices.}
        \label{fig-alpha_pr_pw}
    \end{minipage} \hfill  
    \begin{minipage}{0.24\textwidth}  
        \centering
        \includegraphics[width=\textwidth,height=4.5cm]{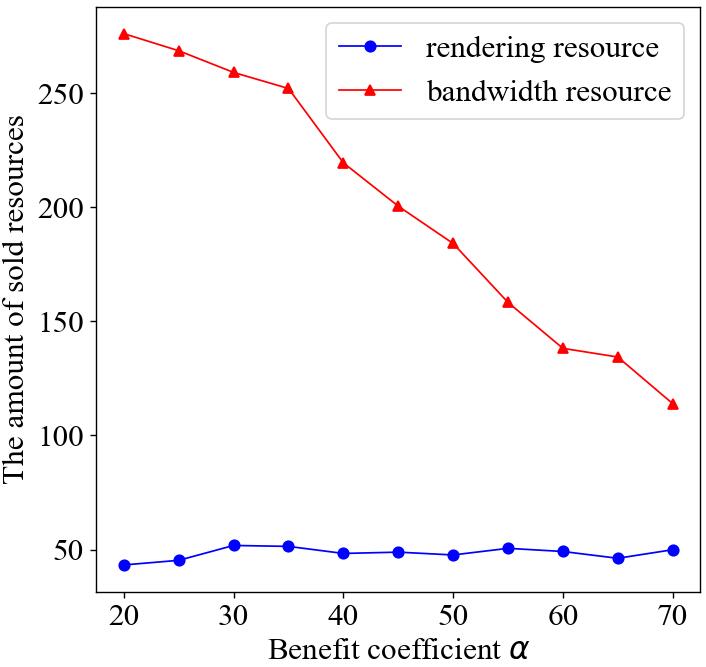}  
        \caption{The effect of $\alpha$ on \quad\quad resource sales.}
        \label{fig-alpha_rendering_bandwidth}
    \end{minipage}
\end{figure*}

 Fig. \ref{fig-user_utility_xir_xiw} depicts the utility of MSU $u_i$ with the fixed unit prices of resources $p_r=p_r^*,\;p_w=p_w^*$. It is obvious that MSU $u_i$ obtain the maximum value of its utility if and only if the purchasing strategy is $({x_i^r}^*=2.32,\;{x_i^w}^*=2.74)$. Similarly, Fig. \ref{fig-MSP_utility_pr_pw} is the profit of the MSP with the fixed purchasing strategies $\{{x_i^r}^*,\;{x_i^w}^*\}_{i=1}^N$. We can find the \quad\quad\quad\quad\quad\quad MSP achieves the maximum utility when choosing the optimal pricing $(p_r^*=40.23,\;p_w^*=22.15)$. The above \quad\quad\quad\quad experiment results indicate that both the MSUs and the MSP cannot change their decisions unilaterally to improve their individual utilities at the Stackelberg Equilibrium (SE) point.

\begin{figure*}[htbp]  
    \centering
    \begin{minipage}{0.24\textwidth}  
        \centering
        \includegraphics[width=\textwidth,height=4.5cm]{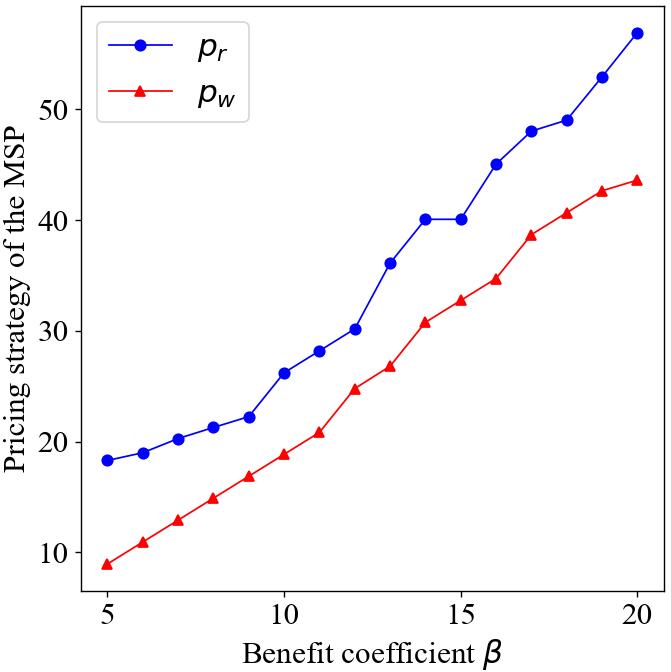}  
        \caption{The effect of $\beta$ on \quad\quad unit prices.}
        \label{fig-beta_pr_pw}
    \end{minipage} \hfill  
    \begin{minipage}{0.24\textwidth}  
        \centering
        \includegraphics[width=\textwidth,height=4.5cm]{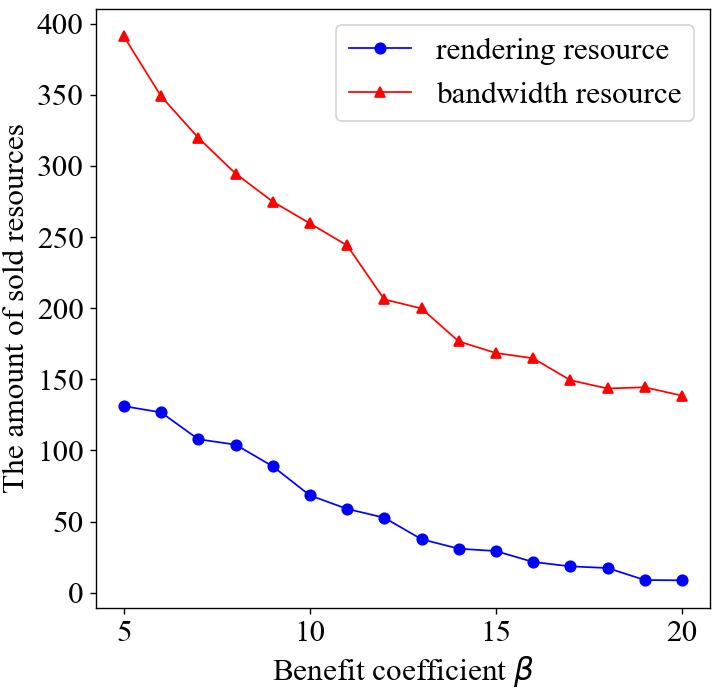}  
        \caption{The effect of $\beta$ on \quad\quad resource sales.}
        \label{fig-beta_rendering_bandwidth}
    \end{minipage} \hfill  
    \begin{minipage}{0.24\textwidth}  
        \centering
        \includegraphics[width=\textwidth,height=4.5cm]{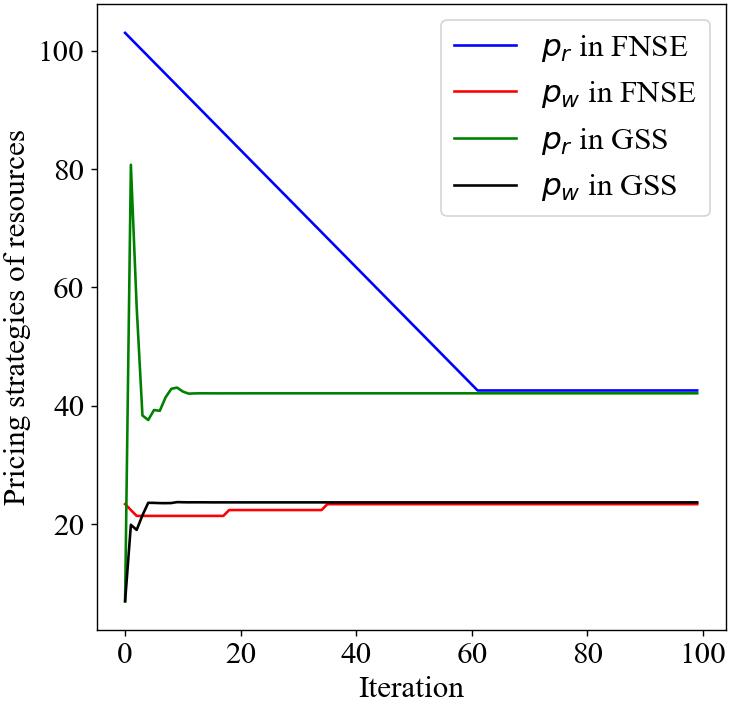}
        \caption{The iterative process of FNSE and GSRAP.}
        \label{fig:FNSE_delta111_kappa111_Golden_pr_pw}
    \end{minipage} \hfill  
    \begin{minipage}{0.24\textwidth}  
        \centering
        \includegraphics[width=\textwidth,height=4.5cm]{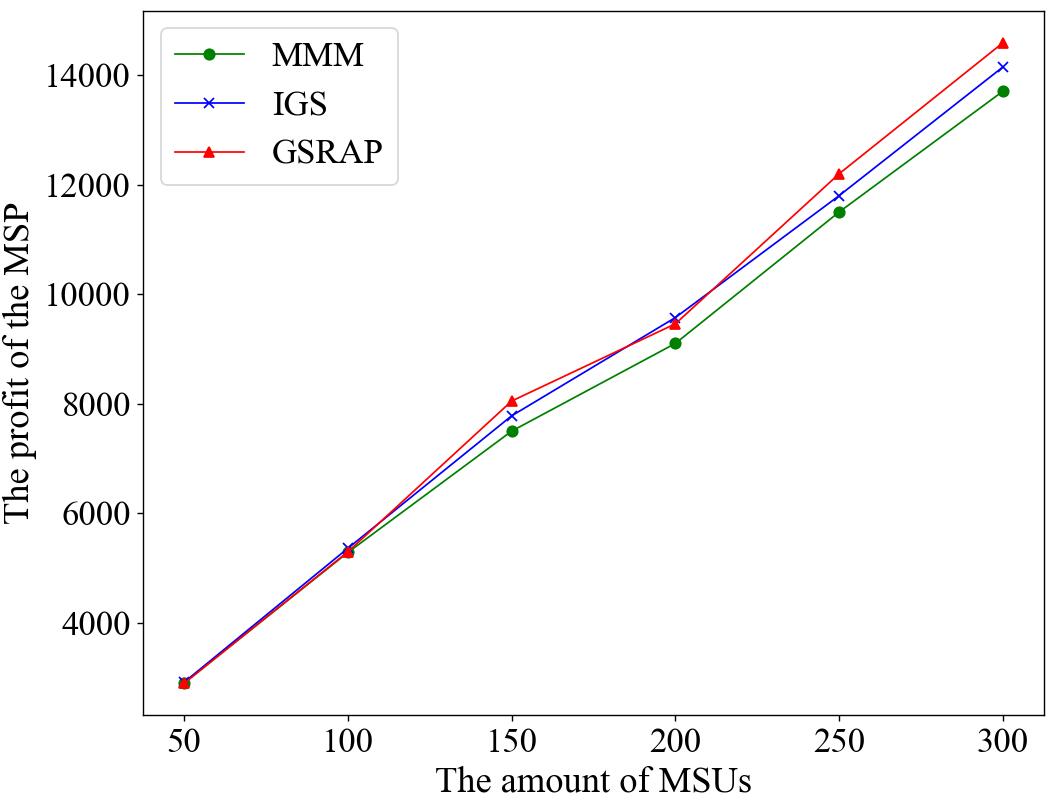}
        \caption{The MSP's profit in \quad different algorithms.}
        \label{fig-algorithms-performance-comparison-MSP-profit}
    \end{minipage}
\end{figure*}

\subsection{Parameter Sensitivity}
\par As depicted in Fig. \ref{fig-alpha_pr_pw}, the price of rendering resources will increase when the benefit coefficient $\alpha$ increases from $20$ to $70$, which is attributed to the fact that users can get more gain from unit rendering resource and allocate more budgets to purchasing rendering resources. Consequently, the MSP will raise the price of rendering resources to get more profits. Besides, we can find that the price of bandwidth rises slowly. This is because bandwidth becomes competitive due to the relatively low price, then bandwidth demands will increase.

\par From Fig. \ref{fig-alpha_rendering_bandwidth}, we can see that the total number of sold bandwidth goes down because the unit price of bandwidth increases. As for the amount of rendering resources, we can find that it rises up at the beginning, then decreases over a period of time, and increases slightly finally. This is because users prefer to allocate more budgets to purchase rendering resources at first. In the medium term, the number of sold rendering resource starts to reduce due to its higher price. Because of the gradually growing price of bandwidth, the MSP can sell more rendering resources at last.

\par In Fig. \ref{fig-beta_pr_pw}, as the parameter $\beta$ increases, the utility provided by one unit bandwidth will be higher, then MSUs are willing to spend more budgets on bandwidth in order to get a more immersive experience. Consequently, the price of bandwidth goes up gradually, which makes rendering resources more competitive than before. We can observe that the price of rendering resources increases gradually from Fig. \ref{fig-beta_pr_pw}. The number of sold rendering and bandwidth resources decreases because the unit prices have raised but MSUs have limited budgets, as shown in Fig. \ref{fig-beta_rendering_bandwidth}.

%
%

\subsection{Performance Comparison}
\par We modify the algorithm FNSE in \cite{Pricing-and-Budget-Allocation-2023} as the baseline algorithm according to our formulated Stackelberg game. Keeping other settings unchanged, we set the initial step $\Delta$ in FNSE to be $1$ in simulations, respectively. The algorithm GSRAP initializes the resource prices as $p_r^{min}$ and $p_w^{min}$ and then search for the optimal solution iteratively.

\par As observed from Fig. \ref{fig:FNSE_delta111_kappa111_Golden_pr_pw}, our algorithm and the baseline algorithm converge to the same Stackelberg equilibrium point finally, which confirms the correctness of our design. Although our algorithm has a relatively large oscillation at first, it can reach the same convergence status faster compared to the algorithm FNSE. In comparison, our algorithm outperforms the baseline algorithm in terms of convergence speed while finding the optimal solution.

\par We compare the performance of different algorithms in the resource allocation process, in which many-to-many matching (MMM) \cite{Computing-Resource-Allocation-Matching-2017} and the algorithm IGS \cite{Cloud-Edge-Computing-2021} are adopted as the baselines. As illustrated in Fig. \ref{fig-algorithms-performance-comparison-MSP-profit}, as the number of MUSs increases, the MSP can obtain more profit because more resource requests will be generated. We can find the MSP's profit in our proposed algorithm GSRAP is higher than two baseline algorithms in most cases. The reason is that our algorithm better considers the balance in the contribution of distinct resources to MSUs' utilities, which prevents one certain resource from being consumed too quickly.

\section{Conclusion}
\par In this paper, we investigate the multi-resource allocation and pricing problem in the metaverse. We first integrate blockchain technology into the metaverse to guarantee the security and reliability of transactions. Subsequently, we formulate the interaction between the MSP and MSUs as a Stackelberg game which takes both rational and irrational MSUs into account. We prove the existence of Stackelberg equilibrium and design an efficient algorithm to find the SE point quickly. Finally, we conduct extensive simulations to validate the correctness and effectiveness of our designs. In the future, we will investigate more complicated multi-leader and multi-follower Stackelberg game models considering the cooperative relationships within leaders.

\end{document}